\documentclass[12pt, reqno]{amsart}
\usepackage{amsmath, amsthm, amscd, amsfonts, amssymb, graphicx, xcolor}
\usepackage[bookmarksnumbered, colorlinks, plainpages]{hyperref}
\usepackage{float}
\theoremstyle{definition}

\theoremstyle{remark}

\numberwithin{equation}{section}

\begin{document}
\setcounter{page}{1}

\color{black}
\noindent 

\centerline{}

\centerline{}

\title[CRBMs in Market Data Generation]{Investigating Conditional Restricted Boltzmann Machines in  Regime Detection}

\author[S. S. Rentala]{Siddhartha Srinivas Rentala}

\address{Master of Science in Quantitative Finance \\ Fordham University}
\email{sr131@fordham.edu} 
\date{\today}

\begin{abstract}
This study investigates the efficacy of Conditional Restricted Boltzmann Machines (CRBMs) for modeling high-dimensional financial time series and detecting systemic risk regimes. We extend the classical application of static Restricted Boltzmann Machines (RBMs) by incorporating autoregressive
conditioning and utilizing Persistent Contrastive Divergence (PCD) to incorporate complex temporal dependency structures. Comparing a discrete Bernoulli-Bernoulli architecture against a continuous Gaussian-Bernoulli variant across a multi-asset dataset spanning 2013–2025, we observe a dichotomy between generative fidelity and regime detection. While the Gaussian CRBM successfully preserves static asset correlations, it exhibits limitations in generating long-range volatility clustering. Thus, we analyze the free energy as a relative negative log-likelihood (surprisal) under a fixed, trained model. We demonstrate that the model’s free energy serves as a robust, regime stability metric. By decomposing the free energy into quadratic (magnitude) and structural (correlation) components, we show that the model can distinguish between pure magnitude shocks and  market regimes. Our findings suggest that the CRBM offers a valuable, interpretable diagnostic tool for monitoring systemic risk, providing a supplemental metric to implied volatility metrics like the VIX.
\end{abstract}

\maketitle


\section{Introduction }

\noindent This study investigates the efficacy of Conditional Restricted Boltzmann Machines (CRBMs) \cite{taylor2006modeling} for modeling high-dimensional financial time series. We extend the classical application of static RBMs \cite{hinton2010practical} by incorporating auto-regressive conditioning and employing Persistent Contrastive Divergence (PCD) to capture complex dependency structures.

Our research studies three challenges in Energy-Based Modeling for finance. First, we aim to overcome the memory limitations of static RBMs by explicitly modeling the transition kernel $P(v_t \mid v_{<t})$, thereby testing the model's ability to capture volatility clustering endogenously while preserving the bipartite graph structure of RBMs. Second, we evaluate the model's capacity to preserve the static correlation structure observed in the market. Third, we investigate whether the free energy of the Boltzmann distribution serves as a robust, unsupervised proxy for systemic market risk and regime detection.

While prior work \cite{kondratyev2019market} relies on thermalization heuristics, this paper provides a rigorous assessment of whether CRBMs can learn a conditional probability density function that respects the memory endemic to financial markets.

\section{Theory}

In this paper, we investigate two types of CRBMs: the first is the Bernoulli-Bernoulli CRBM, and the second is the Gaussian-Bernoulli CRBM. The main difference being that in the former we use a binary discretization to feed visible units $\mathbf{v}$ into our model as opposed to a z-scored continuous variable in the latter. This leads to different mathematical results as shown below. 

\subsection{Energy-Based Formulation}

The Restricted Boltzmann Machine (RBM) is a stochastic neural network rooted in statistical physics\cite{lezmi2020improving}. It defines a joint probability distribution over visible units $\mathbf{v}$ and hidden units $\mathbf{h}$ via the Boltzmann distribution:

\begin{equation}
P(\mathbf{v}, \mathbf{h}) = \frac{1}{Z} e^{-E(\mathbf{v}, \mathbf{h})}
\end{equation}

where $Z = \sum_{\mathbf{v}, \mathbf{h}} e^{-E(\mathbf{v}, \mathbf{h})}$ is the partition function. The partition function acts as a normalization function. By marginalizing out the hidden units, we obtain the probability of the visible vector:

\begin{equation}
P(\mathbf{v}) = \frac{1}{Z} \sum_{\mathbf{h}} e^{-E(\mathbf{v}, \mathbf{h})}
\end{equation}

In this study, we assume the system temperature $T=1$ \cite{li2016temperature}, treating the model as a standard Energy-Based Model (EBM) where learning corresponds to minimizing the energy of observed data samples while maximizing the energy of unobserved configurations\cite{lecun2006tutorial}.

\subsubsection*{The Bernoulli-Bernoulli RBM}
For binary data, the energy function is defined as:

\begin{equation}\label{eq:bernoulli_energy}
E(\mathbf{v}, \mathbf{h}) = - \sum_{i} a_i v_i - \sum_{j} b_j h_j - \sum_{i,j} v_i W_{ij} h_j
\end{equation}

Here, the energy is linear with respect to the states. The binary nature of the units ($v_i, h_j \in \{0, 1\}$) introduces a strong information bottleneck, acting as a regularizer but potentially limiting the model's ability to capture continuous magnitudes, something we prove further in the study. 

\subsubsection*{The Gaussian-Bernoulli RBM}
To model continuous financial returns, we employ the Gaussian-Bernoulli architecture:

\begin{equation} \label{eq:gaussian_energy}
E(\mathbf{v}, \mathbf{h}) = \sum_{i} \frac{(v_i - a_i)^2}{2\sigma_i^2} - \sum_{j} b_j h_j - \sum_{i,j} \frac{v_i}{\sigma_i} W_{ij} h_j
\end{equation}

A critical distinction arises in the first term: $\frac{(v_i - a_i)^2}{2\sigma_i^2}$. Unlike the linear penalty in the Bernoulli case, the Gaussian energy scales quadratically with the distance of the visible unit from its bias. This means that in high-volatility regimes, the quadratic term dominates the energy function, causing the model to assign exponentially lower probabilities to outlier events (large $|v_i|$). Consequently, in high-volatility regimes the free energy of the RBM acts as a distance from the learned equilibrium. In training, this implies that the model may underestimate the probability mass of extreme events (large $|v_i|$). Since these events are assigned disproportionately high energies, they reside on steep gradients of the energy surface. 

\subsection{Derivation of Free Energy}

The central challenge in evaluating the Restricted Boltzmann Machine is the efficient computation of the marginal probability $P(\mathbf{v})$. In a naive formulation, calculating $P(\mathbf{v})$ requires summing over all possible hidden configurations $\mathbf{h}$. For $H$ hidden units, this space has size $2^H$, rendering direct computation intractable\cite{lecun2006tutorial}. However, due to the bipartite structure of the RBM (no intra-layer connections), this summation can be factorized into a linear number of terms.

We define the Free Energy $F(\mathbf{v})$ such that the marginal probability is given by:
\begin{equation}
P(\mathbf{v}) = \frac{e^{-F(\mathbf{v})}}{Z}
\end{equation}
This implies that $F(\mathbf{v}) = -\ln \left( \sum_{\mathbf{h}} e^{-E(\mathbf{v}, \mathbf{h})} \right)$. Below, we derive this function for both binary and continuous architectures.

\subsubsection*{Bernoulli-Bernoulli Free Energy}
For binary inputs, the energy function in summation notation is:
\begin{equation}
E(\mathbf{v}, \mathbf{h}) = - \sum_{i} a_i v_i - \sum_{j} b_j h_j - \sum_{i,j} v_i W_{ij} h_j
\end{equation}
To find the marginal likelihood, we sum the exponentiated negative energy over all binary hidden vectors $\mathbf{h} \in \{0,1\}^H$:
\begin{align}
\sum_{\mathbf{h}} e^{-E(\mathbf{v}, \mathbf{h})} &= \sum_{\mathbf{h}} \exp\left( \sum_{i} a_i v_i + \sum_{j} b_j h_j + \sum_{i,j} v_i W_{ij} h_j \right) \\
&= \exp\left( \sum_{i} a_i v_i \right) \sum_{\mathbf{h}} \exp\left( \sum_{j} h_j \left( b_j + \sum_{i} v_i W_{ij} \right) \right)
\end{align}
Since the hidden units are conditionally independent given $\mathbf{v}$, the exponential of the sum becomes a product of exponentials, allowing us to swap the summation and product operators:
\begin{align}
\sum_{\mathbf{h}} e^{-E(\mathbf{v}, \mathbf{h})} &= \exp\left( \sum_{i} a_i v_i \right) \prod_{j=1}^H \left( \sum_{h_j \in \{0,1\}} \exp\left( h_j \left( b_j + \sum_{i} v_i W_{ij} \right) \right) \right)
\end{align}
The inner sum over $h_j \in \{0,1\}$ has only two terms. If $h_j=0$, the term is $e^0=1$. If $h_j=1$, the term is $\exp(b_j + \sum_i v_i W_{ij})$. Thus:
\begin{equation}
\sum_{\mathbf{h}} e^{-E(\mathbf{v}, \mathbf{h})} = \exp\left( \sum_{i} a_i v_i \right) \prod_{j=1}^H \left( 1 + \exp\left( b_j + \sum_{i} v_i W_{ij} \right) \right)
\end{equation}
Finally, taking the negative logarithm yields the Free Energy:
\begin{equation}
F_{Bernoulli}(\mathbf{v}) = - \sum_{i} a_i v_i - \sum_{j=1}^H \ln \left( 1 + \exp\left( b_j + \sum_{i} v_i W_{ij} \right) \right)
\end{equation}

\subsubsection*{Gaussian-Bernoulli Free Energy}
For continuous financial returns, we substitute the Gaussian energy function. The derivation follows the same logic, but the visible bias term is replaced by the quadratic penalty:
\begin{equation}
E(\mathbf{v}, \mathbf{h}) = \sum_{i} \frac{(v_i - a_i)^2}{2\sigma_i^2} - \sum_{j} h_j \left( b_j + \sum_{i} \frac{v_i}{\sigma_i} W_{ij} \right)
\end{equation}
Factorizing the hidden units as before yields:
\begin{equation}
F_{Gaussian}(\mathbf{v}) = \sum_{i} \frac{(v_i - a_i)^2}{2\sigma_i^2} - \sum_{j=1}^H \ln \left( 1 + \exp\left( b_j + \sum_{i} \frac{v_i}{\sigma_i} W_{ij} \right) \right)
\end{equation}

The summation term $\ln(1 + e^x)$ acts as a \textbf{Softplus} function. Maximizing the probability $P(\mathbf{v})$ is equivalent to minimizing the free energy. This formulation reveals a dynamic interplay between linear and non-linear components depending on the market state:

\begin{enumerate}
    \item \textbf{Feature Extraction (Low Volatility):} While the free energy calculation sums scalar values, the dependency on the input vector $\mathbf{v}$ is non-linear due to the softplus function. During times of low volatility, the quadratic penalty is small, and the free energy is dominated by the second, non-linear term. In this regime, the model minimizes energy by effectively learning the 'valleys' in the energy surface, allowing the hidden units to capture subtle correlation structures and latent factors in 'regular' market days. 

    \item \textbf{Regime Shift (High Volatility):} During times of high volatility, the opposite occurs. The quadratic term $\sum (v_i - a_i)^2 / 2\sigma_i^2$ grows squarely with the magnitude of returns and dominates the energy function. This mathematical behavior mirrors the empirical reality of financial crises, where stable correlations tend to break down. In these moments, the sheer magnitude of idiosyncratic shocks dominates the likelihood of the observed state, effectively overriding the learned correlation structure.
\end{enumerate}

Thus, the free energy serves as a regime-aware metric; it measures structural compliance during calm markets and magnitude deviations during stress.

\subsection{Dynamic Biases in Conditional RBMs}

The primary innovation introduced by the Conditional Restricted Boltzmann Machine (CRBM) \cite{taylor2006modeling} is the conditioning of hidden and visible variables on a fixed window of autoregressed visible nodes, denoted as $v_{<t}$. In this framework, the connections from the past to the current state are \textbf{directed}, ensuring that temporal information flows strictly from past to future ($v_{<t} \to v_t, h_t$).

While the autoregressive lag order can theoretically differ for visible and hidden connections, in this study we set a symmetric lag of $N=5$. This configuration allows the model to capture short-term memory and autocorrelation structures while maintaining the tractability of standard RBM training algorithms. Crucially, by conditioning on the past, the CRBM treats the history as a fixed context, allowing the optimization at time $t$ to proceed as if it were a static RBM with shifted biases.

We define the history vector $v_{<t}$ as the concatenation of the previous $N$ observations. Given $D$ dimensions per time step, $v_{<t}$ has dimension $N \cdot D$. This effectively conditions the system on a fixed set of past variables using two autoregressive weight matrices: $A$ (of dimension $ND \times D$) for the visible units, and $B$ (of dimension $ND \times H$) for the hidden units.

\subsubsection*{Dynamic Hidden Biases}
The static bias $b_j$ is replaced by a time-dependent dynamic bias $\hat{b}_{j,t}$, which incorporates the weighted contribution from recent past observations:

\begin{equation}
\hat{b}_{j,t} = b_j + \sum_{k=1}^{ND} B_{kj} v_{k,<t}
\end{equation}

This dynamic bias effectively modulates the activation threshold of each hidden node $j$ according to the history of the market. Intuitively, as the model trains over batches, $B_{kj}$ learns to shift the effective prior of the hidden units based on trends observed in $v_{<t}$.

In the model this means that for $h_t$ we first initialize the hidden node based on the lagged effects on the state and then train on $v_t$, effectively setting a prior. This is what theoretically allows, given a suitable $N$, to observe volatility clustering effects in the synthetic data.

To elucidate further, this mechanism ensures persistence in regime detection. For example, if the market crashes (-10\%) at time $t$, the autoregressive component (Matrix $B$) significantly increases the dynamic bias $\hat{b}_{j,t+1}$ for the 'crisis-detecting' hidden units. Even if the subsequent move at $t+1$ is smaller (e.g., -1\%), the elevated bias ensures that the "Crisis Regime" remains the \textbf{energy minimum} (most probable state). Effectively, Matrix $B$ dynamically reshapes the energy landscape, digging a "valley" for the high-volatility state to ensure the model does not prematurely revert to the mean.

Consequently, the probability of a hidden unit being active is governed by both this dynamic bias and the current input $v_t$:

\begin{equation}
P(h_{j,t} = 1 \mid v_t, v_{<t}) = \frac{1}{1 + \exp\left(-\hat{b}_{j,t} - \sum_{i} W_{ij} v_{i,t}\right)}
\end{equation}

\subsubsection*{Dynamic Visible Biases and Reconstruction}
Similarly, the visible units receive a dynamic bias $\hat{a}_{i,t}$, which is an affine function of the past observations. This term explicitly models the conditional mean of the time series:

\begin{equation}
\hat{a}_{i,t} = a_i + \sum_{k=1}^{ND} A_{ki} v_{k,<t}
\end{equation}

The reconstruction of the visible units at time $t$, given the hidden state and history, is modeled as a Gaussian distribution centered on the sum of the dynamic visible bias and the feedback from the hidden layer:

\begin{equation}
P(v_{i,t} \mid h_t, v_{<t}) = \mathcal{N}\left(\hat{a}_{i,t} + \sum_{j} W_{ij} h_{j,t}, \; 1\right)
\end{equation}

This formulation allows the CRBM to distinguish between structure that is predictable from the past (captured by $A$ and $B$) and the innovations or "surprises" at time $t$ (captured by the interaction weights $W$). Note that in the binary case this probability takes a similar form to the hidden bias inference. 

\section{Data and Methodology}

This study utilizes a multi-asset dataset spanning from January 2013 to January 2025, curated to capture a diverse range of market factors including Equities, Rates, Volatility, and Credit spreads. The specific assets were chosen to ensure the model is exposed to both risk-on/risk-off dynamics and structural correlations across asset classes. The complete asset universe, detailed in Table \ref{tab:asset_universe}, includes key indicators such as the S\&P 500 for equity distributions, the VIX for implied volatility, and the 10Y-2Y Treasury spread for recessionary signaling.

\begin{table}[h]
\centering
\resizebox{\textwidth}{!}{%
\begin{tabular}{|l|l|l|}
\hline
\textbf{Ticker} & \textbf{Category} & \textbf{Descriptor} \\
\hline
SPX Index & Equities & S\&P 500 Daily Return (Target variable distribution) \\
VIX Index & Equities & Implied Volatility (VIX) ("Fear" gauge) \\
VIX3M Index & Equities & 3-Month Implied Volatility (Term structure) \\
SKEW Index & Equities & S\&P 500 Skew Index (Tail risk demand) \\
MTUM US Equity & Equities / Factor & iShares US Momentum Factor ETF \\
VLUE US Equity & Equities / Factor & iShares US Value Factor ETF \\
QUAL US Equity & Equities / Factor & iShares US Quality Factor ETF \\
USGG2YR Index & Rates & 2-Year Treasury Yield (Near-term expectations) \\
USGG10YR Index & Rates & 10-Year Treasury Yield (Long-term expectations) \\
USYC2Y10 Index & Rates & 10Y–2Y Spread (Yield curve slope) \\
MOVE Index & Rates / Volatility & ICE BofA MOVE Index (Bond volatility) \\
LUACOAS Index & Credit & Investment Grade Corporate OAS \\
LF98TRUU Index & Credit & High Yield Corporate OAS \\
DXY Index & FX / Macro & US Dollar Index (Global liquidity proxy) \\
CL1 Comdty & Commodities & WTI Crude Oil Front Month \\
GC1 Comdty & Commodities & Gold Front Month (Safe-haven) \\
CESIUSD Index & Macro & Citi Economic Surprise Index \\
USGGBE10 Index & Macro & 10Y Breakeven Inflation Rate \\
BFCIUS Index & Macro & Bloomberg US Financial Conditions Index \\
\hline
\end{tabular}%
}
\caption{Asset Universe and Descriptors}
\label{tab:asset_universe}
\end{table}

To evaluate the efficacy of different CRBM architectures, we employ two distinct preprocessing pipelines. Method A (Discrete) follows the methodology of Kondratyev (2019)\cite{kondratyev2019market} with the added CRBM, where continuous returns are discretized into binary vectors using a 16-bit encoding scheme; this allows the Bernoulli-Bernoulli CRBM to process continuous market data while maintaining high granularity through $2^{16}$ distinct bins. In contrast, Method B (Continuous) employs Z-Score Standardization for the Gaussian-Bernoulli CRBM, where data is normalized to zero mean and unit variance ($v_{i,t} = (x_{i,t} - \mu_i)/\sigma_i$). This normalization is critical for the Gaussian energy function, as the quadratic penalty assumes inputs fall within a standard range to ensure stable gradients\cite{taylor2006modeling}.

The model's generalization capability is assessed through a chronological split that separates distinct market regimes. The Training Set encompasses the period from 2013 to 2019, representing a relatively stable, low-volatility bull market that allows the model to learn baseline correlations. The Testing Set, covering 2020 to 2025, introduces extreme structural breaks, including the COVID-19 crash, the subsequent liquidity-fueled rally, and the 2022 inflation regime. This  split specifically tests the Conditional RBM's ability to utilize its dynamic bias mechanism to adapt to unseen high-volatility environments that differ fundamentally from the training distribution.

\section{Results and Analysis}

The results and analysis section is divided into a shorter section on the Bernoulli-Bernoulli CRBM and a slightly more verbose analysis on the Gaussian-Bernoulli CRBM. Note that all analysis for model selection was done on the basis of mean squared error as suggested by Hinton \cite{hinton2010practical}. 

\subsection{Bernoulli-Bernoulli CRBMs} 
We evaluate the generative fidelity of the Bernoulli-Bernoulli CRBM by examining the Quantile-Quantile (QQ) plots of real versus synthetic returns. As illustrated in the left panel, the real market data, as expected (Post-2020), exhibits severe leptokurtosis, with extreme quantiles.

\begin{figure}[H]
    \centering
    \includegraphics[width=1.0\linewidth]{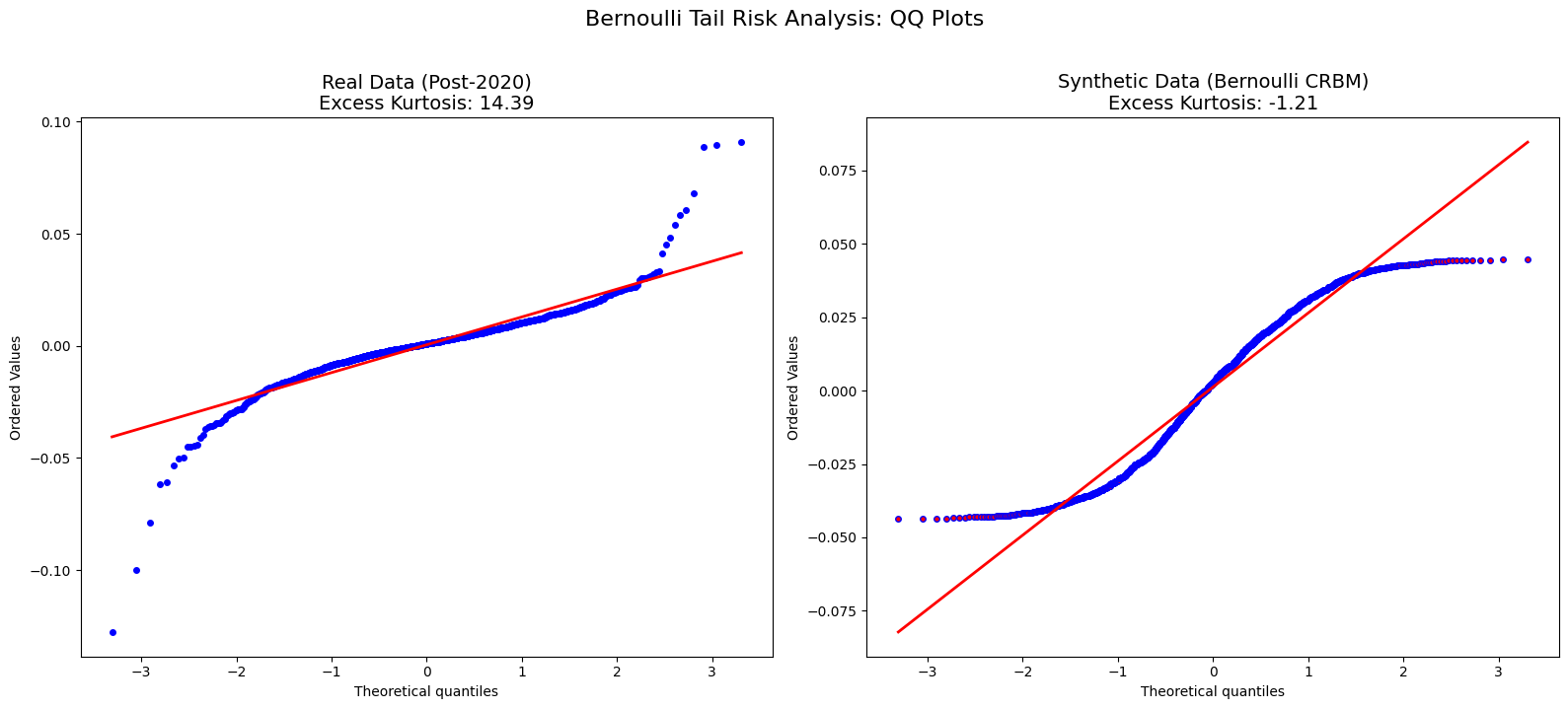}
    \caption{QQ Plot of Real vs Synthetic Data (Bernoulli-Bernoulli)}
    \label{fig:qqplot_bernoulli}
\end{figure}

In contrast, the synthetic data generated by the Bernoulli model (right panel) demonstrates a distinct \textbf{undercounting of tail risks}. The synthetic quantiles plateau at the extremes, effectively truncating the distribution and failing to reproduce the magnitude of outlier events. This can be attributed to two structural constraints inherent in the Bernoulli-Bernoulli architecture:

\begin{enumerate}
    \item \textbf{Bit Flipping:} The limiting factor of our model is not resolution, but the structure of the binary encoding. Standard binary representation introduces 'Hamming Cliffs' where small changes in continuous value require simultaneous flips of multiple bits. This non-smooth optimization landscape hinders the gradient descent process, preventing the model from learning subtle correlations.
    \item \textbf{The Binary Bottleneck:} Unlike continuous models that learn manifolds in $\mathbb{R}^N$, the Bernoulli RBM must model dependencies between binary bits. This creates a severe information bottleneck. The model tends to learn the robust correlations of the center of the distribution (which it fits well) while treating the rare, active bits in the tail bins as noise. This then 'averages out' the volatile data points, thus resulting in thin tails. 
\end{enumerate}

Correspondingly, the correlation matrix of the synthetic data exhibits a severe \textbf{dampening of inter-asset correlations} (Figure 2) 

This failure can be attributed to the structural limitations of the Bernoulli architecture. By decomposing continuous returns into discrete bits, the model destroys the inherent covariance structure of the data. To recover these correlations, the weight matrix $\mathbf{W}$ must learn highly specific dependencies between the \textbf{individual bits of different assets}, something we attempt to rectify using the Most Significant Bit methodology. Nonetheless, this imposes an excessive representational burden that the training process failed to resolve, confirming that the binary bottleneck is a fundamental impediment to modeling multi-asset dependence.

\begin{figure}[H]
    \centering
    \includegraphics[width=1.0\linewidth]{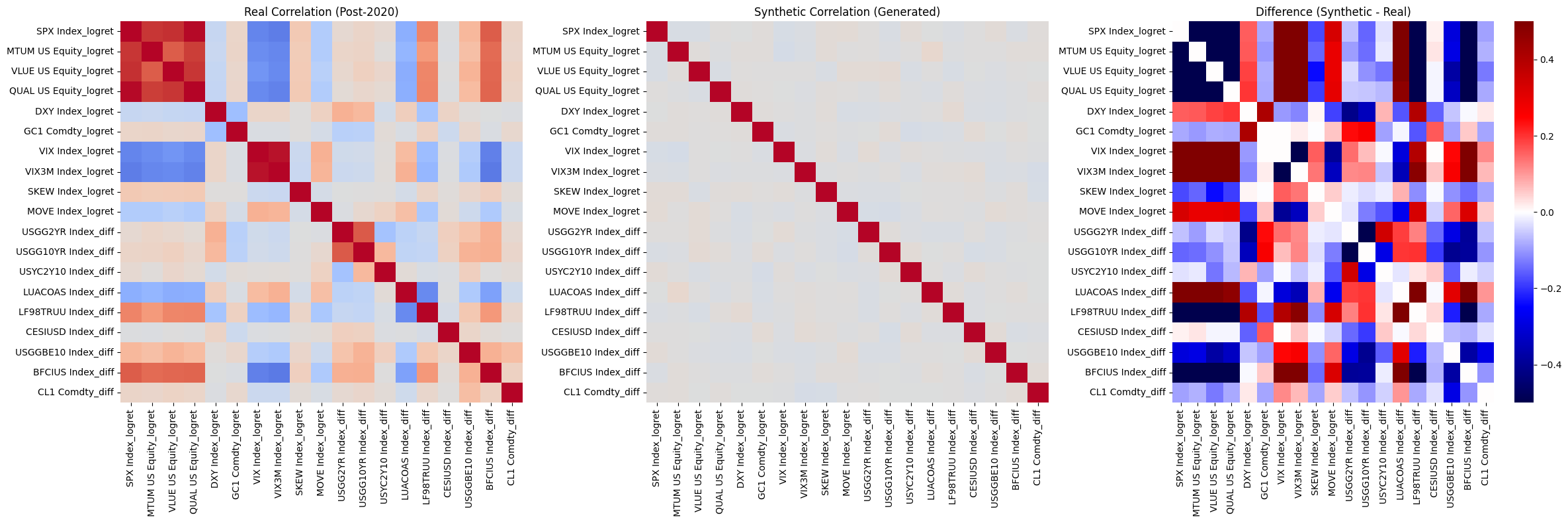}
    \caption{Correlation Plot of Real and Synthetic Data (Bernoulli-Bernoulli Model)}
    \label{fig:corr_bernoulli}
\end{figure}
\begin{figure}[H]
    \centering
    \includegraphics[width=0.5\linewidth]{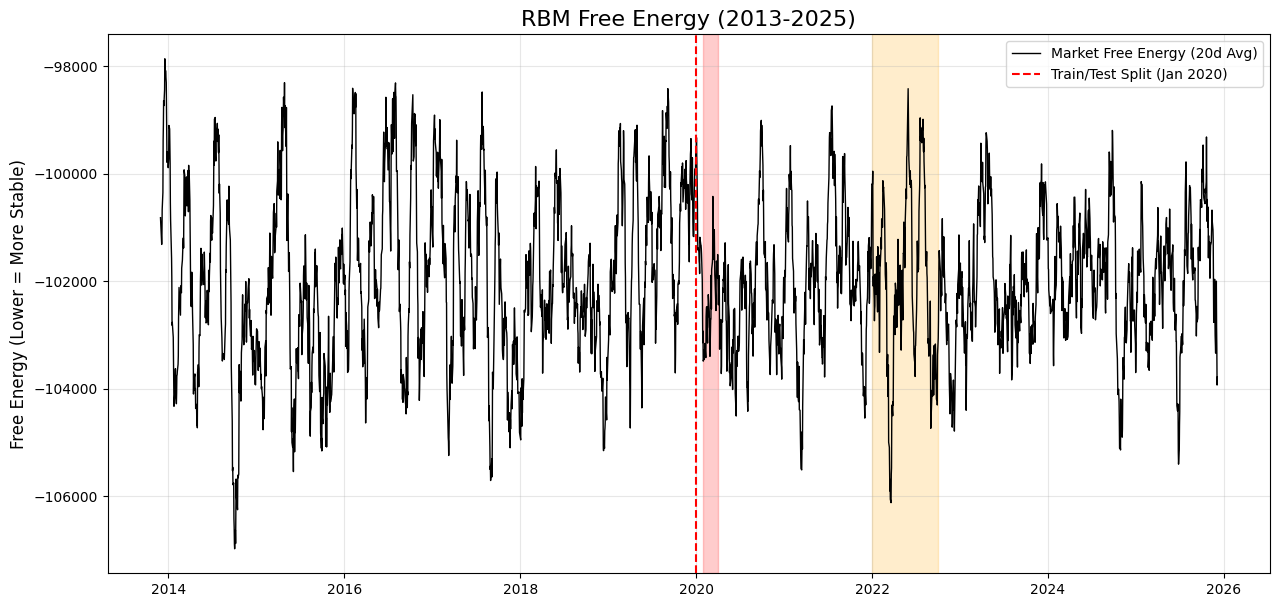}
    \caption{Bernoulli Free Energy Plot}
    \label{fig:RBM Free Energy plot}
\end{figure}

Finally, we examine the trajectory of the free energy $F(\mathbf{v})$ over the 2013-2025 period to determine its utility as an \textbf{unsupervised metric for systemic risk}, something first introduced in Section 2.2. Theoretically, a robust CRBM should assign lower energy to stable market states and higher energy to unseen, volatile regimes. However, as illustrated in Figure 3, the free energy signal is dominated by high-frequency noise, rendering it ineffective for regime detection.

Instead of clear, discernible shifts during stress periods (such as the COVID-19 crash in early 2020 or the 2022 inflation shock), the free energy exhibits mean-reverting oscillations. This instability is, once again, a direct artifact of the bit flipping we elucidated upon earlier. In a 16-bit binary space, minor fluctuations in continuous asset returns can trigger rapid "bit-flipping". These changes in the visible vector $\mathbf{v}$ create stochastic noise that drowns out the lower-frequency structural signal of the market. While the MSB priority should, in theory, rectify this by attributing the left-most bits as those that the model should pay more attention to, this doesn't hold simply due to the amount of noise introduced in the remaining flipped 15 bits for every data point. 

Furthermore, the absolute magnitude of the free energy raises concerns regarding the model's discriminative capability. The values consistently hover in the extreme range. Since $P(\mathbf{v}) \propto e^{-F(\mathbf{v})}$, these deeply negative energies imply that the model assigns near-certain probabilities to effectively all observed states, something we are unable to fully rectify even with sparsity targets as suggested by Hinton \cite{hinton2010practical}. The model fails to differentiate between high-probability normal market days and low-probability crisis days; instead, it collapses the probability landscape, treating every state as highly likely. Consequently, the Bernoulli formulation lacks both the smoothness and the contrast required to serve as a reliable market stability metric. 

\subsection{Gaussian-Bernoulli CRBM Results and Analysis} 
We evaluate the generative fidelity and the overall market stability implied in the Gaussian-Bernoulli CRBM. The former, we do by examining the QQ plots of real vs synthetic returns. As depicted in the figure below, we can clearly observe that the synthetic data collapses data into a gaussian distribution. The model clearly does a good job of fitting the central probability mass but fails at the tails, we posit that this is because the quadratic term in the energy function of the Gaussian-Bernoulli function, in the extreme outliers, effectively forces the $P(\mathbf{v})$ to a standard normal distribution, thus meaning that we, by definition, fail to learn the tails of the distribution. 
\begin{figure}[H]
    \centering
    \includegraphics[width=1.0\linewidth]{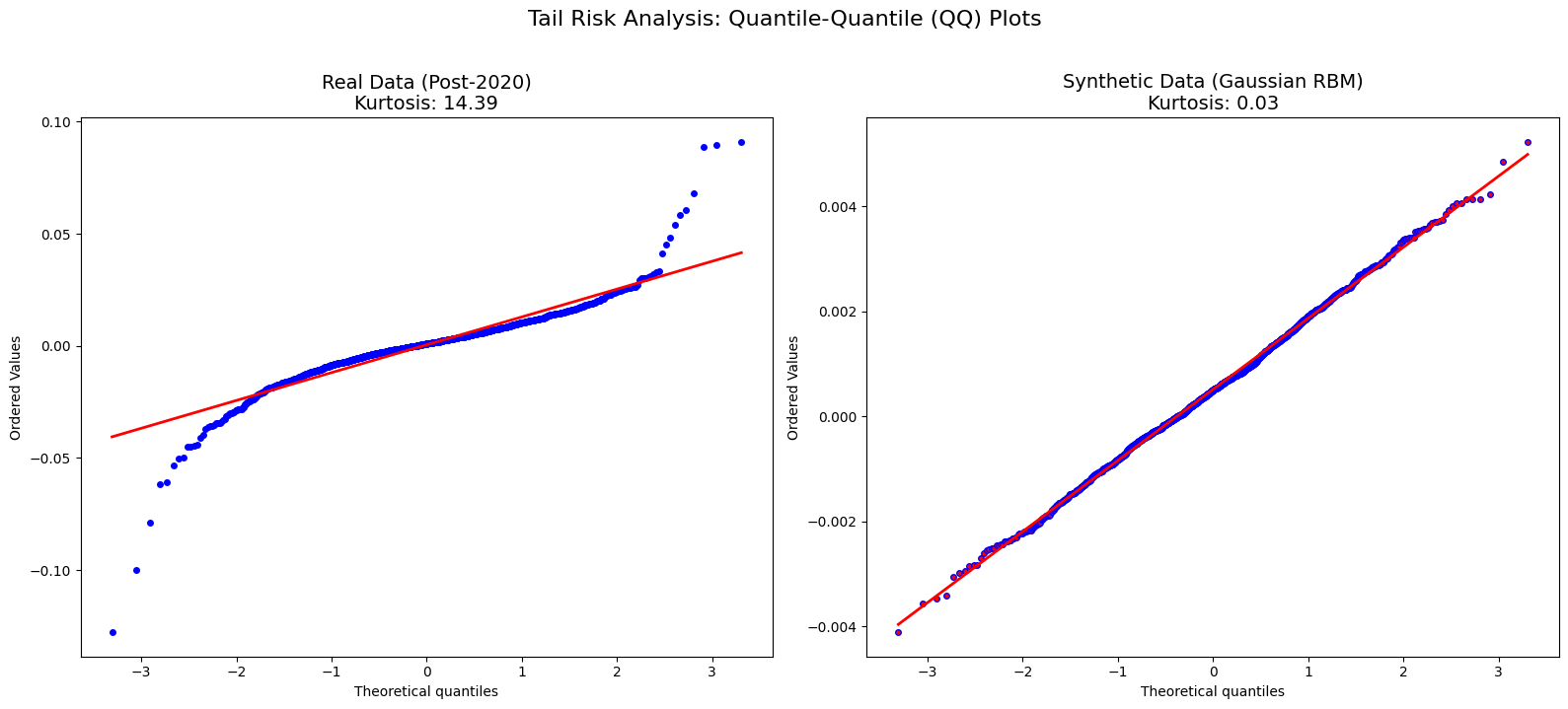}
    \caption{QQ Plot of Real vs Synthetic Data (Gaussian-Bernoulli)}
    \label{fig:qq_plot_gaussian}
\end{figure}

Conversely, the correlation structure of the assets does very well, as depicted below. Indeed, we can see that on average, the model performs extremely well in comparison to our previous model. This is attributable to the following:
\begin{enumerate}
    \item \textbf{Autoregressive Conditioning (Matrices $\mathbf{A}$ and $\mathbf{B}$):} The inclusion of directed autoregressive connections allows the model to explicitly learn temporal dependencies. By conditioning the hidden state on the past ($v_{<t}$), the model captures the persistence of correlation regimes. Even with a fixed lag of $N=5$, this memory mechanism enables the model to sustain complex dependency structures that static RBMs fail to retain. 
    \item \textbf{Continuous Latent Space:} The shift from 16-bit quantization to continuous Gaussian inputs ($\mathbf{v} \in \mathbb{R}^D$) eliminates the information loss caused earlier. Rather than forcing the model to learn relationships between independent bits, the Gaussian units allow the weights $\mathbf{W}$ to directly encode the covariance magnitude between assets.
\end{enumerate}
\begin{figure}[H]
    \centering
    \includegraphics[width=1.0\linewidth]{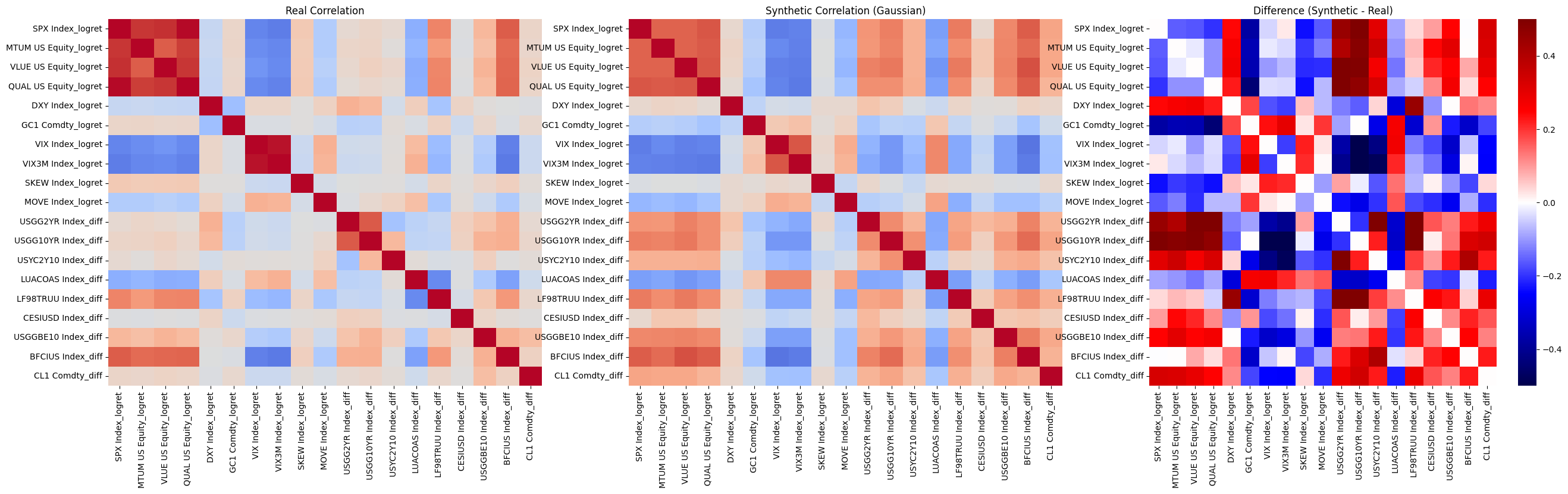}
    \caption{Correlation Structure of Synthetic and Real Data (Gaussian-Bernoulli)}
    \label{fig:gaussian_corr}
\end{figure}

\subsubsection{The Free Energy as a proxy for market stability}

The final result of this paper explores the use of free energy (defined in Section 2.2) as a proxy for the stability of the market. Mathematically, we have shown that high free energy is a result of improper alignment of weights, biases, and visible nodes. Additionally, the quadratic term would dominate the free energy term; resulting in the trained model associating low probability to that set of inputs. Below we overlay the free energy of the model over the VIX series.  

\begin{figure}[H]
    \centering
    \includegraphics[width=1.0\linewidth]{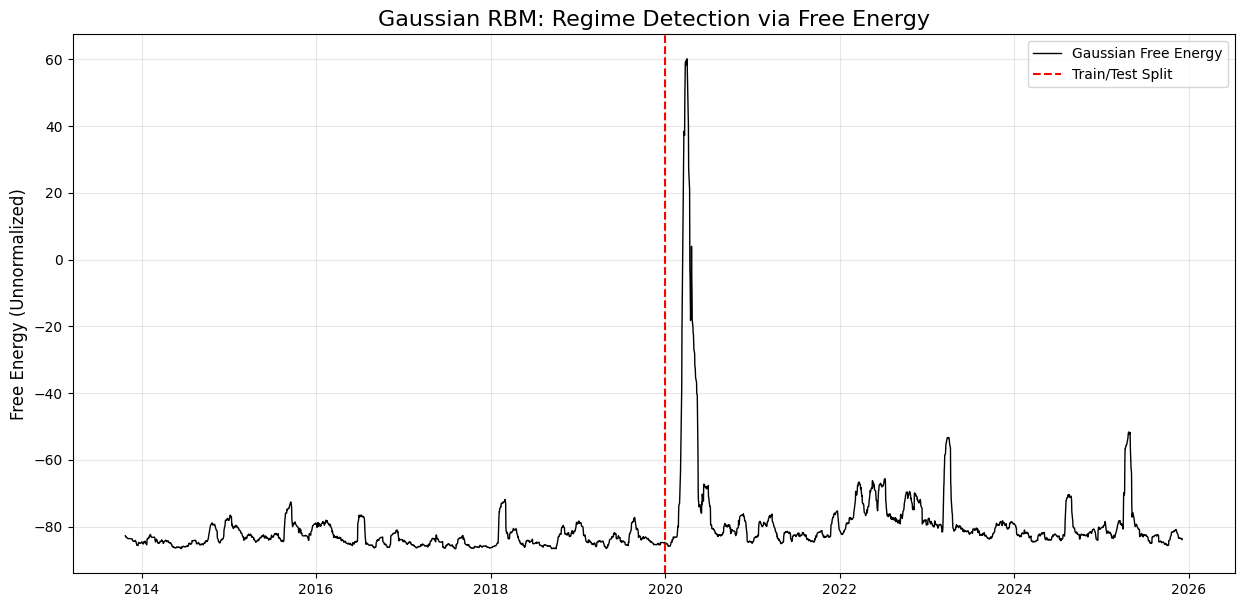}
    \caption{Free Energy Over Time}
    \label{fig:gaussian_fe}
\end{figure}
\begin{figure}[H]
    \centering
    \includegraphics[width=1.0\linewidth]{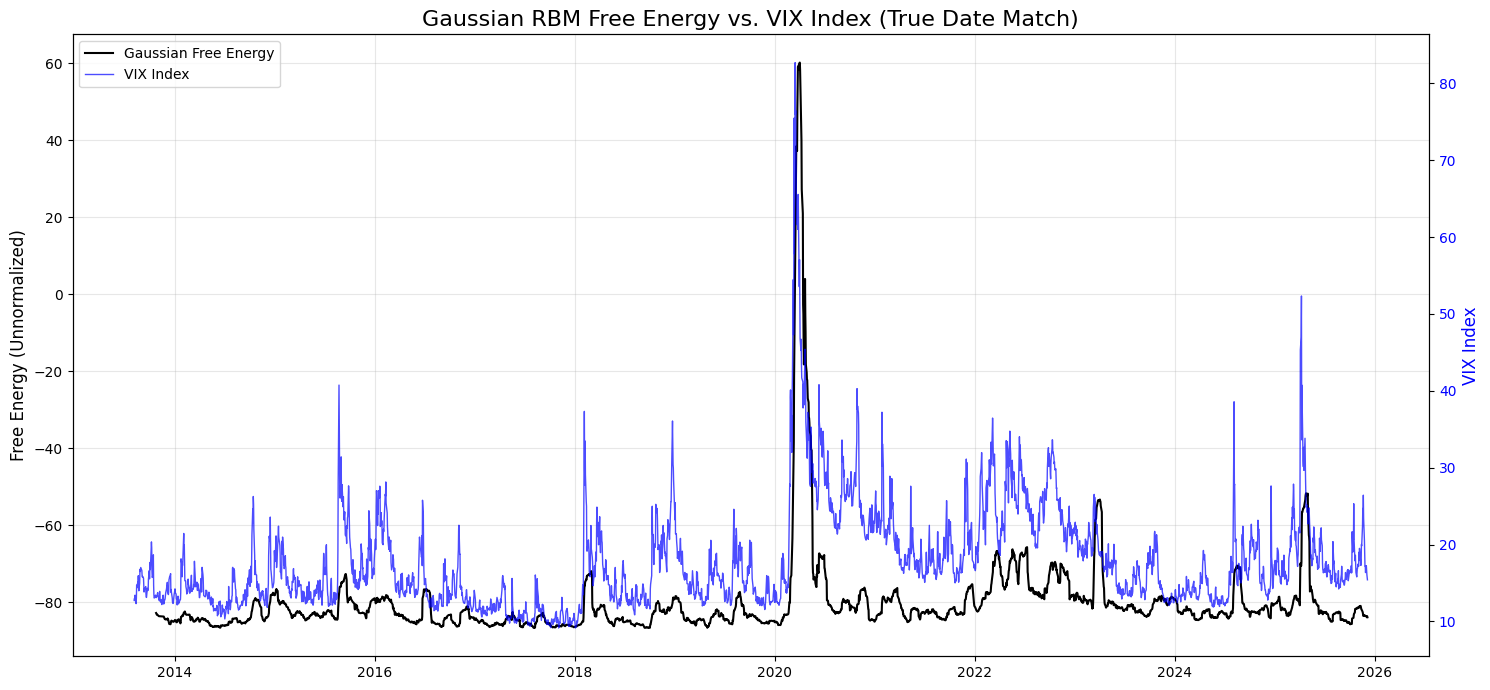}
    \caption{Free Energy and VIX Index over time}
    \label{fig:fe_vix}
\end{figure}
We can clearly observe that the free energy spikes during market crashes (as the model assigns low probabilities to these out-of-sample states). Indeed, we can also see that the model, even in-sample, does a relatively decent job of discriminating between known periods of high volatility such as 2018 and other calmer market states. The massive spike observed during the COVID-19 market crash (out of sample), showcases the extent to which the model is 'surprised' by such a crash, and the fact that this energy does not immediately fall to the baseline. Nonetheless, the VIX remains elevated even after a crash, while the free energy indicates that our model suggests correlations have approached a level of normalcy. Nonetheless, an understanding of the markets implies that this is a counterintuitive result. Furthermore, we would expect that the observed post-2020 free energy of the market would settle at a higher baseline. While this is somewhat true in the model, we cannot make a concrete claim here. 

To verify that the free energy is not merely a proxy for the $L_2$ norm of returns (the quadratic term), we decompose the metric into its constituent parts: the Quadratic Reconstruction Term (measuring magnitude) and the Structural Softplus Term (measuring alignment with learned features). By plotting these against the VIX, we can isolate the contribution of learned correlations to the anomaly detection signal.

\begin{figure}[H]
    \centering
    \includegraphics[width=1.0\linewidth]{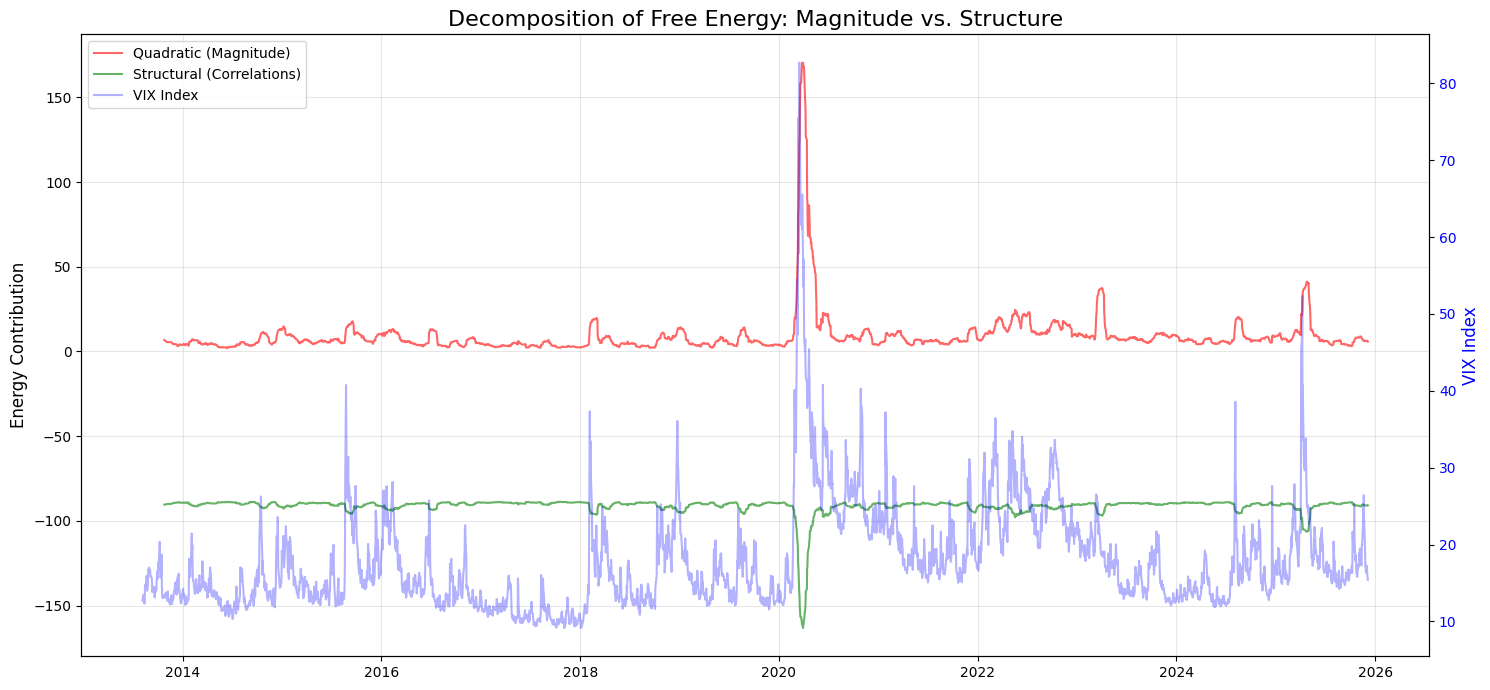}
    \caption{Quadratic and Structural Terms}
    \label{fig:fe_decomposition}
\end{figure}

This decomposition reveals a clear  separation in how the Gaussian-Bernoulli CRBM processes market stress. The quadratic term ($\sum (v_i - a_i)^2 / 2\sigma_i^2$) acts as the realized volatility, showcasing a massive spike during the 2020 crash that perfectly tracks the VIX; this is expected behavior, as the term scales quadratically with the squared deviation of standardized returns. During times of market normalcy, the quadratic term is very close to zero and has minimal contribution to the total free energy; suggesting that these are very high probability states according to the model. 

However, the true insight lies in the structural term ($-\sum \ln(1 + e^{b_j + \mathbf{v}^T \mathbf{W}_{\cdot j}})$). During the same crash period, this term drops sharply, becoming significantly more negative.  the dot product between the learned weight vector $\mathbf{W}$ and the input vector $\mathbf{v}$ explodes during the crash, the activation of these specific hidden units drives the structural energy down. This inverse relationship between the quadratic term and the structural term confirms that the model does not view the crash merely as high-variance noise, but rather as a distinct, recognized correlation structure; a specific "crisis regime" that aligns strongly with the model's learned representation. Conversely, the structural term approaches its 'maximum' during other periods; suggesting that the model believes that during these periods the crisis nodes are switched off. 

The true result from the above analysis is that the VIX by itself, often overestimates risk and corresponds to the quadratic term even though at a base level, we observe that market correlations have not shifted significantly (as depicted by the structural term). Essentially, the model suggests that if the structural term is flat and the VIX spikes, the VIX may be over-counting the risk whereas the free energy captures both dimensions. 
\section{Conclusion and Future Work}

This paper examined Conditional Restricted Boltzmann Machines as statistical energy-based models for high-dimensional financial data, with the goal of assessing their usefulness for unsupervised regime detection rather than as physically grounded market models. By comparing a discretized Bernoulli–Bernoulli architecture with a continuous Gaussian–Bernoulli formulation, we isolated the practical consequences of model class assumptions on both generative performance and diagnostic behavior.

The Bernoulli CRBM was shown to suffer from structural limitations inherent to binary representations, including information bottlenecks and unstable likelihood estimates driven by bit-level noise. In contrast, the Gaussian CRBM preserved the empirical correlation structure of the market and produced a stable likelihood signal over time, albeit at the cost of thin-tailed synthetic generation due to its quadratic term. This limitation is not a training failure but a direct consequence of the assumed energy function.

Interpreting the CRBM strictly as a likelihood model, we analyzed its free energy as a relative surprisal measure under a fixed parameterization, rather than as a thermodynamic quantity. Decomposing this surprisal into magnitude-dependent and interaction-dependent components revealed that periods of market stress are not homogeneous: some episodes are dominated by large return realizations with relatively stable dependence structure, while others coincide with pronounced deviations in the model’s learned cross-asset features. This distinction cannot be captured by volatility measures alone and highlights the potential diagnostic value of free energy based approaches in regime analysis.

Several limitations must be emphasized. The model assumes conditional independence across time given a finite lag window and does not possess an internal latent state capable of generating true dynamical persistence. Furthermore, fixing the variance of the visible units forces all heteroscedastic effects to be absorbed through bias shifts, contributing to elevated surprisal during extreme events. Finally, the use of Gaussian visible units precludes the accurate modeling of heavy-tailed return distributions.

\subsection{Future Directions}
To address the limitations identified in this study, future research should focus on the following key areas:
\begin{enumerate}
    \item Heavy-Tailed Energy Functions: The Gaussian assumption inherently penalizes outliers quadratically, forcing the model to prioritize the bulk of the distribution during training. Replacing the Gaussian visible units with Student-t units or using a Huber Loss energy function would allow the model to assign higher probability to tail events without destabilizing the learning process, potentially resolving the ”thin tail” generative failure, though it may lead to further difficulty in convergence.
    \item Deep Belief Networks (DBNs): The current shallow architecture limits the model’s ability to learn hierarchical features (e.g., ”Industry factor stress” → ”Market Crash”). Stacking multiple RBMs into a Deep Belief Network could allow higher layers to capture slower-moving, abstract
    regime variables, while lower layers handle the high-frequency noise of daily returns. 

\end{enumerate}
In conclusion, while the CRBM may not yet replace standard econometric models for path generation, its ability to unsupervisedly learn and detect structural regime shifts offers a valuable, interpretable tool for systemic risk monitoring in modern financial markets.

\providecommand{\bysame}{\leavevmode\hbox to3em{\hrulefill}\thinspace}
\providecommand{\MR}{\relax\ifhmode\unskip\space\fi MR }
\providecommand{\MRhref}[2]{%
  \href{http://www.ams.org/mathscinet-getitem?mr=#1}{#2}
}

\end{document}